\documentclass{XrU2005}
\usepackage{graphicx}
\usepackage{natbib}
\newcommand{\dd}{$\rm deg^{2}$}
\newcommand{\flux}{$\rm erg\,cm^{-2}s^{-1}$}

\begin{document}
\title{The XMM-LSS cluster sample and its cosmological applications
}
\author{M. Pierre, F. Pacaud}
{ \affil{DAPNIA/SAp CEA Saclay, 91191 Gif sur Yvette, France}
\author{the XMM-LSS consortium}
{\affil{http://vela.astro.ulg.ac.be/themes/spatial/xmm/LSS/cons\_e.html}
\maketitle

{\sl Talk given at ``The X-ray Universe 2005"\\
ESA Symposium\\
 El Escorial, 26-30 September 2005}

\begin{abstract}
We present the  X-ray source detection procedure that we have
developed for the purpose of assembling and characterizing
controlled samples of cluster of galaxies for the XMM Large Scale
Structure Survey. We describe how we model the selection function
by means of simulations: this leads us to define source classes
rather than flux limited samples. Focussing on the CFHTLS D1 area,
our compilation suggests a cluster density higher than previously
determined from the deep ROSAT surveys above a flux of $2\times
10^{-14}$ \flux. We also present the L-T relation for the 9
brightest objects in the area. The slope is in good agreement with
the local correlation. The relation shows luminosity enhancement
for some of the $0.15 < z< 0.35$ objects having $1 < T <2 $ keV, a
population that the XMM-LSS is for the first time systematically
unveiling.
\end{abstract}
\section{Introduction}
The question of cosmic structure formation is substantially more
complicated than the study of the spherical collapse of a pure
dark matter perturbation in an expanding Universe. While it is
theoretically possible to predict how the shape of the
inflationary fluctuation spectrum evolves until recombination, we
hardly understand the subsequent galaxy, AGN and cluster formation
because of the problems of non-linear growth and the feedback from
star formation. Consequently we cannot use the statistics of
visible matter fluctuations to constrain the nature of the Dark
Matter and Dark Energy without developing this
understanding of non-gravitational processes. \\
Clusters, the most massive entities of the Universe, are a crucial
link in the chain of understanding. They lie at the nodes of the
cosmic network, and have virialized cores, but are still growing
by accretion along filaments. The rate at which clusters form, and
the evolution of their space distribution, depends strongly on the
shape and normalization of the initial power spectrum, as well as
on the Dark Energy equation of state (e.g. \citet{rapetti05}).
Consequently, both a 3D mapping of the cluster distribution and an
evolutionary model relating cluster observables to cluster masses
and shapes (predicted by theory for the average cluster
population)  are needed to test the consistency of the ``CMB WMAP
concordance cosmology" with the properties of clusters in the
low-$z$ Universe.
With its mosaic of 10 ks overlapping XMM pointings, the XMM Large
Scale Structure  survey (XMM-LSS) has been designed to detect a
significant fraction of the galaxy cluster population out to a
redshift of unity over an area of several tens of square degrees,
so as to constitute a sample suitable to address the questions
outlined above \citep{pierre04}. The trade-off in the survey design
was depth versus
coverage, keeping within reasonable total observing times. \\
This configuration allows investigation of the cluster population
down to a depth of about $ 10^{-14}$ \flux\ which is comparable to
the deepest ROSAT serendipitous surveys \citep{rosati02}. However,
observations are performed with a  narrower PSF (FWHM $\sim$ 6"
for XMM vs $\sim$ 20 " for the ROSAT PSPC) and very different
different instrumental characteristics such as background noise
and focal plane configuration. This led us to develop a new  X-ray
pipeline which is presented in the next section, along with the
principles of the  computation of the survey selection function.
Spectroscopically identified clusters undergo detailed
measurements, which enabled us to track the evolution of the
low-mass end of the cluster population (Sec. 3). Source statistics
and the L-T relation for the CFHTLS D1 area are
presented in Sec. 4.\\
In the following, we assume a $\Lambda$CDM cosmology and give all fluxes
in the  [0.5-2] keV band.
\section{{\tt Xamin} -- A new X-ray pipeline}
\subsection{Design}
The two major requirements of the XMM-LSS X-ray processing were to
reach the sensitivity limit of the data in a  statistically
tractable manner in terms of cluster detection efficiency, and hence to
calculate the selection function of the detected
objects. The package that we have developed, {\tt Xamin}, combines
the sensitivity of the multi-resolution wavelet analysis for
source detection with the rigour of a likelihood analysis for
assessing the significance of the detected sources, handling the
complex XMM instrumental characteristics. Both steps use
Poisson statistics. The whole procedure has been validated by
means of extensive simulations of point-like and extended sources
\citep{pacaud05}.
\subsection{Source classification}
Simulations of the LogN-LogS X-ray point source population  give
for {\tt Xamin} a 90\% completeness level  down to a flux of
$4\times 10^{-15}$ \flux\ for 10 ks exposures. As a rule of thumb,
the corresponding sensitivity for `typical' cluster
sources is 2-3 times higher. However, cluster detection
efficiency depends not only on the object flux and size, and the
instrumental PSF,  but also very much on the background level and
on the detector topology such as CCD gaps and vignetting, as well
on the ability of the pipeline to separate close pairs of
pointlike sources.  We thus stress that the concept of sky
coverage, i.e. the fraction of the survey area covered at a given
flux limit, is strictly valid only for point-sources because, for
faint extended objects, the detection efficiency is surface
brightness limited (rather than flux limited). Moreover, since
the faint end of the cluster luminosity function is poorly
characterised at $z>0$,
it is not possible to estimate  a posteriori what   fraction of
groups   remain undetected, unless a cosmological model is
assumed, along with a thorough modelling of the cluster
population out to high redshift.\\
Consequently, with the goal of constructing deep controlled
samples suitable for cosmology   we define, rather than   flux
limits, two classes of extended sources corresponding to specific
levels of contamination and completeness. The selection is
performed   in the {\tt Xamin} output parameter space defined by
{\tt extent likelihood,  extent, detection likelihood}. Selection
criteria have been established by means of extensive simulations
for cluster apparent core-radii (extent) ranging from 10" to 100",
and total number counts from 50 to 1000. The cluster surface
brightness distribution was assumed to follow a $\beta$-model,
with $\beta = 2/3$. The C1 class is defined to have ``no
contamination". i.e. no point sources missclassified as extended.
For the C2 class, selection criteria are relaxed such as to allow
for 50\% contamination by spurious extended sources. The
classification has been in turned checked against some 60
spectroscopically XMM-LSS clusters confirmed to date. C1 clusters are high
surface brightness extended objects; this selection inevitably
retains a few  nearby galaxies, but these are readily discarded
from the sample by inspection of optical overlays. The C2 sample
includes fainter clusters than C1, and also a number of
nearby galaxies, saturated point sources and unresolved pairs, as
well as cases badly effected by CCD gaps; the contamination is a
posteriori removed by the visual inspection of the optical
overlays as well as by the outcome of the spectroscopic
identification programme. The principle of the procedure is
illustrated in Fig. \ref{class}.
\begin{figure}
\centering  \includegraphics[width=8cm]{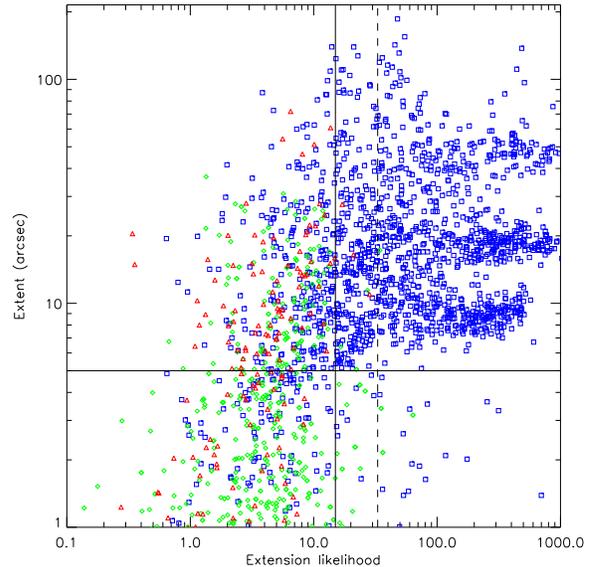}
\caption{ Cluster selection criteria in the {\tt Xamin} parameter
space populated by simulations. Point sources are represented by
green diamonds, clusters by blue squares; red triangles indicate
spurious sources. The horizontal and vertical solid lines
delineate the C1+C2 class. The dotted vertical line indicates the
restriction to the C1 class to which the {\tt detection
likelihood} $> 32$ condition was added, guaranteing an
uncontaminated sub-sample of clusters \label{class}}
\end{figure}
\begin{figure}
\centering \includegraphics[width=8cm]{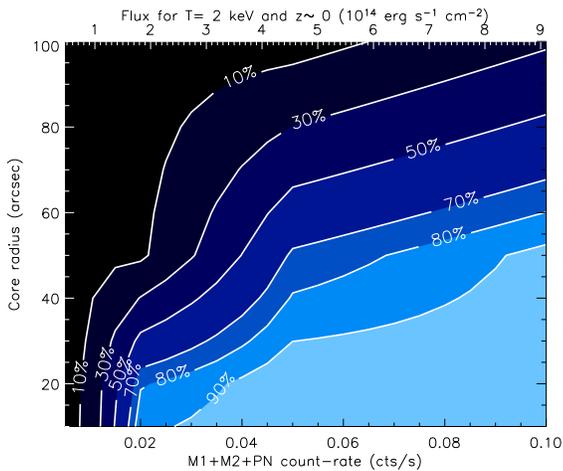} \caption{
Detection probability for C1-type clusters as a function of
count rate and core-radius, averaged for the 20 inner arcmin of the
XMM field, and for an exposure time of  10 ks  \label{probC1}}
\end{figure}
\subsection{The survey selection function}
In parallel, simulations provide the necessary basis for the
computation of the selection function. They allow us to derive
detection probabilities as a function of source core-radius and
countrate for any exposure time, background level and position on
the detector (Fig. \ref{probC1}).  As an example, we show  the
$dn/dz$ prediction  for the C1 cluster population, assuming the
following model: $\Lambda$CDM cosmology \citep{bennett03} + P(k)
power spectrum with transfer function from \citet{bardeen86} and
 comoving halo number density from the
\citet{sheth99} mass function, $M_{200}-T$ relation from
\citet{arnaud05}, $L-T$ relation from \citet{arnaud99} and a
constant core radius of 180 kpc. The cluster/group population was
simulated  down to $T$ = 1~keV (no evolution of the cluster M-T-L
scaling laws was assumed, as it is currently unconstrained by
observations for the $1<T<3$ keV range which constitute the bulk
of our population ). The predicted C1 $dn/dz$ is shown in Fig.
\ref{dndz}, along with the observed redshift distribution of the
observed C1 population. The agreement is very satisfactory and the
data suggest a deficit of clusters around a redshift of 0.5,
probably induced by a cosmic void. Observed cluster numbers from
the currently available 51 XMM-LSS pointings are 7/\dd\ and
12/\dd\ for the C1 and C2 class respectively.
\begin{figure}
\centering \includegraphics[width=8cm]{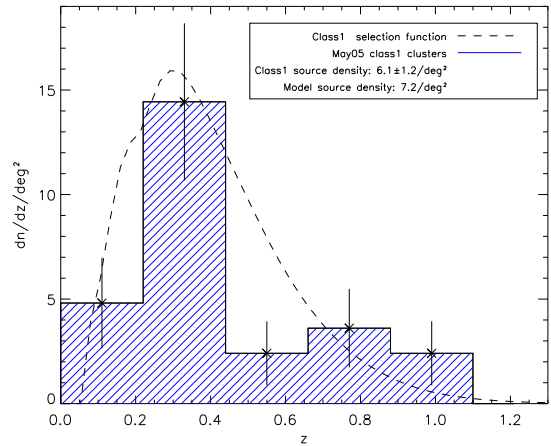} \caption{
Redshift distribution of the C1 clusters for the current XMM-LSS
area: 48 pointings covering  4.6 effective \dd\ (only the inner 20
arcmin are considered) -- that is 29 objects. The dotted line
corresponds to the predictions from a simple halo model in a
$\Lambda$CDM cosmology (see text).
 \label{dndz}}
\end{figure}
Finally, we have investigated for our simple cosmological model,
to what extent the C1 and C2 classes are comparable to flux
limited samples (Fig. \ref{lost}). The main result is that the C1
sample approaches, in terms of number density,  a flux limited
sample of about $ 2\times 10^{-14}$ \flux. But (1) more high-z
objects are detected, while nearby low-surface brightness groups
are not retained, and (2) the C1 sample is strictly defined from
X-ray criteria (and hence is not contaminated) while it would be necessary
to examine a large number  of sources to clean a   putative flux
limited sample at  $ 2\times 10^{-14}$ \flux, with many of them not
being unambiguously characterized as extended or point-like.
\begin{figure}
\centering
\includegraphics[width=8cm]{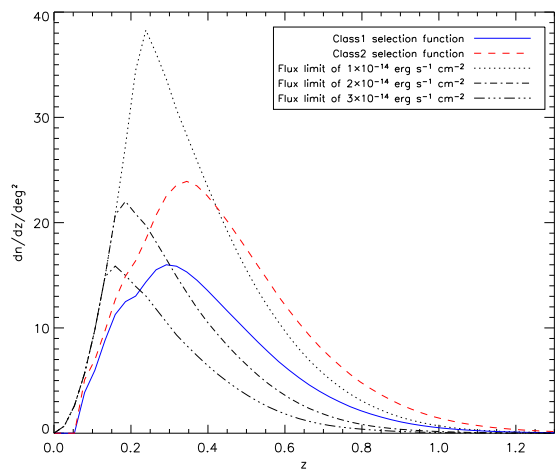}\caption{Comparison of the
C1 and C1+C2 redshift distribution with various flux-limited
samples, assuming the simple cosmological model described in the
text
 \label{lost}}
\end{figure}
\section{Cluster measurements}
Each spectroscopically confirmed cluster  undergoes  detailed
spectral and spatial  fits in order to determine its temperature,
flux  and bolometric luminosity within $R_{500}$ (standard
overdensity radius). In particular, we have demonstrated that,
under specific statistical model and binning conditions, we reach
a 20\% accuracy in temperature measurements with $\sim 100$ and
$\sim 300$ photons for 1 and 2 keV groups respectively
\citep{willis05}. It turns out that the C1 cluster sample is
almost identical to the sample for which we can measure
temperatures.
\section{Results from the D1 CFHTLS area}
The D1 CFHTLS area covers 1\dd\ and constitutes the central part
of the XMM-LSS (see \citet{pierre04} for a general layout of
multi-wavelength coverage associated with the XMM-LSS). It
includes among others the Vimos VLT Deep Survey (VVDS,
\citet{bondi03}). We present a summary of the properties of the
cluster sample for this region\footnote{XMM exposure time is 20 ks
over the region but we keep the same class definition as the S/N
increase is only $\sqrt 2$; the net effect being only a slight
increase in C1/C2 number density  ratio for this sub-region}. We
have spectroscopically identified   13 clusters over the 0.8 \dd\
effective area. Out of these, 8 are C1 clusters, 1 is classified
as C2, and 4 are clusters not entering the classification (very
faint objects or clusters contaminated by a point source). The
selection function for these 4 latter systems is unknown, but they
are interesting objects indicative of our ultimate detection
limit, i.e. $\sim 3-5\times 10^{-15}$ \flux. Two examples are
illustrated on Figs. \ref{C1} and \ref{C3}.
\begin{figure}
\centering \includegraphics[width=8cm]{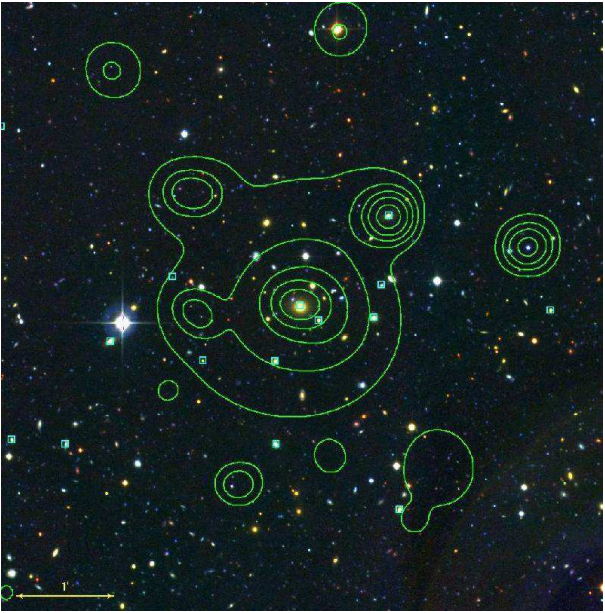}
\caption{Example of a C1 cluster at z = 0.26 (F $\sim 8\times
10^{-14}$ \flux). The cluster X-ray contours are overlaid on a
u,r,z CFHTLS composite. They are drawn from the co-added [0.5-2]
keV MOS1+MOS2+pn mosaic, filtered in the wavelet space using a
significance threshold of $10^{-3}$ for Poisson statistics (not
corrected for vignetting). The intensity scale is logarithmic
(counts/pixel/second, not corrected for vignetting). Boxes indicate
cluster members for which we obtained a redshift using LDSS at the
Magellan telescope.
 \label{C1}}
\end{figure}
\begin{figure}
 \centering \includegraphics[width=8cm]{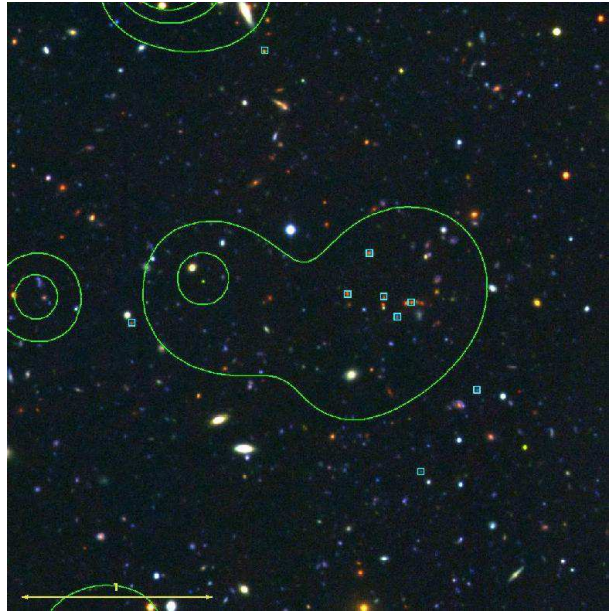} \caption{ Example of a faint
 cluster at z = 0.92 at the detection limit (F $\sim 3\times 10^{-15}$ \flux).
The source was not spectroscopically followed-up. Redshift
confirmation comes from the VVDS data alone. Same symbols and
contours as in Fig. \ref{C1}
 \label{C3}}
\end{figure}
The C1+C2 clusters span the $0.05<z<1.05$ redshift range, with
bolometric luminosities ranging from 0.1  to $4\times10^{44}$
erg\,s$^{-1}$.
\\
7 clusters are found above a flux of $ 2\times 10^{-14}$ \flux.
This translates to about   8.5/\dd.
\begin{figure}
\centering
\includegraphics[width=8cm]{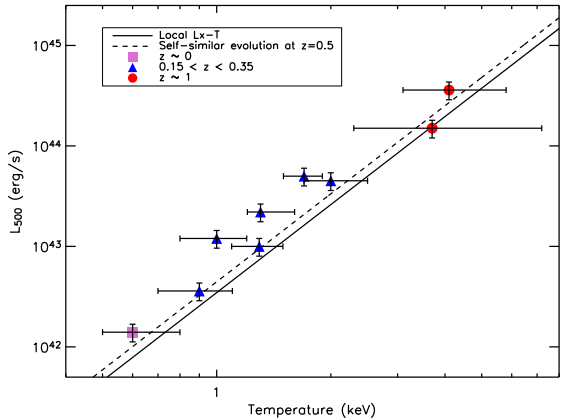}
\caption{  L($R_{500}$)-T relation for the D1 area clusters for
which a temperature was obtained.
      The solid lines gives the mean local L-T relation, while the dotted
      line is the expected luminosity
      enhancement from the self-similar model, at a mean redshift of 0.5.
      The various symbols show the 3 distinct
      redshift (and luminosity/mass) regimes of the measured clusters.
 \label{LT}}
\end{figure}
We have investigated the L-T relation for the 9 clusters for which
it was possible to measure a temperature. Results are displayed on
Fig. \ref{LT} as a function of redshift. Malmquist bias is here
obvious since a temperature was derived only for the apparently
brightest clusters.  The overall correlation appears to be quite
tight and well summarizes the ability of rather shallow XMM
survey-type observations to provide important new insights into
the cluster population. This is particularly relevant for groups
out to $z\sim0.4$ that the XMM-LSS is systematically unveiling for
the first time.  Our intermediate redshift subsample of groups
(0.9-2 keV for $z_{med} = 0.25$) contains objects more luminous
than predicted by the self similar evolution model. According to
this hypothesis, the luminosity scales as the Hubble constant if
it is integrated within a radius corresponding to a fixed ratio
with respect to the critical density of the universe
\citep{voit05}. From the local universe,  we know that low
temperature groups show a larger dispersion in the L-T relation
than massive clusters \citep{helsdon03}. This reflects their
individual formation histories, since they are particularly
affected by non-gravitational effects, as well as the possible
contributions from their member galaxies. The apparent biasing
toward more luminous objects and/or cooler system could come from
the fact that we detect more easily objects having a central cusp,
i.e. putative cool-core groups. This is under investigation using
the full sample available from the current XMM-LSS area (Pacaud et
al in preparation). Detailed presentation of the D1 catalogue and
results are given by \citet{pierre05}.

\section{Summary and conclusions}
We have developed an X-ray pipeline  enabling  the construction,
from shallow XMM pointings, of samples of galaxy clusters  with
controlled selection effects. In particular, ``first class"
clusters constitute a sample selected   upon X-ray criteria alone
(once nearby galaxies are removed).  This approach offers the
advantage of avoiding hypotheses about  the faint end of the
cluster LogN-LogS (currently not explored beyond the local
universe) since this can become critical for faint samples said to
be ``flux limited''. The C1 class contains objects as faint as
$\sim 10^{-14}$ \flux\  out to redshifts of unity. The final C1+C2
sample reaches a density of 12/\dd\ but the initial C2 selection
has to be cleaned (by inspection of overlays and possibly optical
spectroscopy) of
a similar density of spurious cluster candidates ($\sim $5/\dd).\\
We systematically unveil the low end ($T< 2$ keV) of the cluster
population out to a redshift of 0.4.
For the D1 CFHTLS area, we find a cluster density of 8.5 clusters
per \dd\ having a flux larger than $2 \times 10^{-14}$ \flux. This
is higher than the 4-5 clusters /\dd\ given by the RDCS LogN-LogS
\citep{rosati98} and the shallow XMM/2dF survey
\citep{gaga05}, which finds 7/2.3 = 3 /\dd\ for the same flux limit.
Given the size of the studied area, our results are certainly
subject to   cosmic variance which  may also affect the 2.3 \dd\
XMM/2df survey and, to a lesser extent, the  RDCS, which covers 5 \dd\
at $2 \times 10^{-14}$ \flux\ \citep{rosati02}. For comparison,
our simple cosmological model predicts some 7.5 clusters /\dd\
having $T>1 $ keV above a flux of $2 \times 10^{-14}$ \flux\
\citep{pacaud05}. \\
We present the first L-T relation from XMM survey-type
observations over a contiguous area.  The relation contains 9
clusters out to a redshift of unity  over only $\sim 1$\dd\ and
so, opens remarkable perspectives for the study the evolution of the
scaling laws for the cluster/group population with moderate XMM
exposure times.
On-going work aims at characterizing  the evolution of the  L-T
relation for the $T<2$~keV population over the currently available 7
\dd\ of the XMM-LSS (including the Subaru Deep Survey). In
parallel, extensive simulations of XMM images enable us to
investigate the effect of the cluster scaling laws (M-L-T-R), and
of their evolution, on the survey selection function. These will
be further constrained by the APEX and AMiBA Sunyaev-Zel'dovich
surveys of the region to be performed in 2006. It will then be
possible to model in a self-consistent way the cluster population
in parallel with constraining cosmological parameters.
\section*{Acknowledgments}
This work is based on data obtained with XMM,   VLT/FORS2,
VLT/VIMOS, NTT/EMMI, Magellan/LDSS, and imaging campaigns
performed at CTIO and CFHT\footnote{The cluster optical images
were obtained with MegaPrime/MegaCam, a joint project of CFHT and
CEA/DAPNIA, at the Canada-France-Hawaii Telescope (CFHT) which is
operated by the National Research Council (NRC) of Canada, the
Institut National des Science de l'Univers of the Centre National
de la Recherche Scientifique (CNRS) of France, and the University
of Hawaii. This work is based in part on data products produced at
TERAPIX and the Canadian Astronomy Data Centre as part of the
Canada-France-Hawaii Telescope Legacy Survey, a collaborative
project of NRC and CNRS}.
\bibliographystyle{aa}
\bibliography{mmp_d1}
\end{document}